# Subjecting a graphene monolayer to tension and compression

Georgia Tsoukleri[1,2], John Parthenios[1,2], Konstantinos Papagelis[3], Rashid Jalil[4], Andrea C. Ferrari[5], A. K. Geim[4], Kostya S. Novoselov[4] and Costas Galiotis[1,2,3]

[1]*Institute of Chemical Engineering and High Temperature Chemical Processes, Foundation of Research and Technology-Hellas (FORTH/ICE-HT), Patras, Greece*

[2]*Interdepartmental Programme in Polymer Science and Technology, University of Patras, Patras, Greece*

[3]*Materials Science Department, University of Patras, Patras, Greece*

[4]*Department of Physics and Astronomy, Manchester University, Manchester, UK*

[5]*Engineering Department, Cambridge University, Cambridge, UK*

**The mechanical behaviour of graphene flakes under both tension and compression is examined using a cantilever-beam arrangement. Two different sets of samples were employed involving flakes just supported on a plastic bar but also embedded within the plastic substrate. By monitoring the shift of the 2D Raman line with strain, information on the stress transfer efficiency as a function of stress sign and monolayer support were obtained. In tension, the embedded flake seems to sustain strains up to 1.3%, whereas in compression there is an indication of flake buckling at about 0.7% strain. The retainment of such a high critical buckling strain confirms the relative high flexural rigidity of the embedded monolayer.**

The mechanical strength and stiffness of crystalline materials are normally governed by the strength and stiffness of their interatomic bonds. In brittle materials, defects present at the microscale are responsible for the severe reduction of tensile strengths from those predicted theoretically. However, as the loaded volume of a given brittle material is reduced and the number of microscopic defects diminishes, the material strength approaches the intrinsic – molecular- strength. This effect was first described by Griffith in 1921[1] and the best manifestation of its validity is the manufacture and use of thin glass and carbon fibres that nowadays reinforce a whole variety of commercial plastic products such as sports goods, boats, aircrafts, etc.

With reference to material stiffness, the presence of defects plays a minor role and is rather the degree of order and molecular orientation that provide the amount of stiffness along a given axis. In other words, in order to exploit the high stiffness in crystals, the stress direction should coincide with the eigenvector of a given bond.[2] Pure stretching of covalent or ionic bonds is normally responsible for high material stiffness whereas bending or twisting provides high compliance. This is why commercial- amorphous- polymers are compliant materials since an external stress is mainly consumed in the unfolding of entropic macromolecular chains rather than stretching of individual bonds.[2]

Graphene is a two-dimensional crystal, consisting of hexagonally-arranged covalently bonded carbon atoms and is the template for one dimensional CNTs, three dimensional graphite, and also of important commercial products, such as polycrystalline carbon fibres (CF). As a single, virtually defect-free crystal, graphene is predicted to have an intrinsic tensile strength higher than any other known materials[3] and tensile stiffness similar to graphite.[4] Recent experiments [4], have confirmed the extreme tensile strength of graphene of 130 GPa and the similar in-plane Young's modulus of graphene and graphite, of about 1TPa.[4] One way to assess how effective a material is in the uptake of applied stress or strain along a given axis is to probe the variation of phonon frequencies upon loading. Raman spectroscopy has proven very successful in monitoring phonons of a whole range of materials under uniaxial stress[5] or hydrostatic pressure.[6] In general, phonon softening is observed under tensile loading and phonon hardening under compressive loading or hydrostatic pressure. In graphitic materials, such as CF[7] the variation of phonon frequency per unit of strain can provide information on the efficiency of stress transfer to individual bonds. This is because when a macroscopic stress is applied to a polycrystalline CF, the resulting deformation emanates not only from bond stretching or contraction (reversible molecular deformation), but also from a number of other mechanisms such as crystallite rotation and slippage, which do not change the phonon frequency.[5] Indeed, the higher the crystallinity of a fibre (and hence the modulus) the higher the degree of bond deformation and, hence, the higher the measured Raman shift per unit strain.[8]

The recently developed method for graphene preparation by micromechanical cleavage of graphite[9] provides an opportunity for studying the Raman band shifts of both G and 2D modes[10] upon tensile or compressive loading at the molecular level.[11]-[14] This is important not only for highlighting the extreme strength and stiffness of graphene but also to link its behaviour with the mechanical deformation of other graphitic structures such as bulk graphite, CNT and CF. The G peak corresponds to the doubly degenerate $E_{2g}$ phonon at the Brillouin zone centre. The D peak is due to the breathing modes of $sp^2$ rings and requires a defect for its activation.[10],[15] It comes from TO phonons around the **K** point of the Brillouin zone[10],[15], is active by double resonance[16] and is strongly dispersive with excitation energy due to a Kohn Anomaly at **K.**[17] The 2D peak is the second order of the D peak. This is a single peak in monolayer graphene, whereas it splits in four in bilayer graphene, reflecting the evolution of the band structure.[10] Since the 2D peak originates from a process where momentum conservation is obtained by the participation of two phonons with opposite wavevectors it does not require the presence of defects for its activation, and is thus always present. Indeed, high quality graphene shows the G, 2D peaks, but not D.[10]

The first measurement of 2D peak variation with applied strain in a high modulus poly(acrylonitrile) (PAN) derived CF was reported in Ref. [18]. It was recently shown that the 2D peak has a large variation with strain in graphene $\partial\omega_{2D}/\partial\varepsilon \sim 64$ cm$^{-1}$/%[13], where $\omega_{2D}$ is the position of the 2D peak, Pos(2D), and $\varepsilon$ the applied uniaxial strain. Refs [11],[12],[14] have also measured the 2D variation as a function of applied tension in graphene, but reported significantly lower values than in [13]. The Raman scattering geometry used for the case of PAN-based CF which have "onion-skin morphology" (re: large multi-wall nanotubes) [7],[8],[18] is analogous to that of graphene and bulk graphite.[13] Hence, a comparison between the strain sensitivity in tension for all 3 classes obtained by different groups can be attempted, as shown in Table 1. The results for graphene obtained by different authors can differ by a factor of 2 or more. Furthermore, some values reported for graphene[11],[12],[14], are similar to those measured on fibres[18], which we consider fortuitous in view of the polycrystalline nature of the fibres of Ref.[18].

In all past work reported above, stress was transferred to graphene by the flexure of plastic substrates[11]-[14]. However, the adhesion forces between the exfoliated flakes and the polymer molecules are of van der Waals nature, which may not be of sufficient magnitude to (a) transfer stress to graphene and (b) restrain it from slippage during flexure. In Ref. [13] we have applied the strain very slowly over three bending and unbending cycles and used two different set-ups. We took the consistency of the data and the excellent agreement of the Gruneisen parameter measured for the G peak with that reported in hydrostatic experiments on graphite as evidence of no slippage. In Ref. [14] narrow strips of titanium were deposited on the sample in order to clamp it on the substrate, but the measured shifts are still much smaller than Ref. [13]. Refs. [11], [12] just assumed no slippage and, hence, did not take particular steps to minimise it.

In this work, we set out to perform mechanical experiments on graphene sheets employing poly(methyl methacrylate) (PMMA) cantilever beams.[19] As explained in the Experimental section, the advantage of this approach over other conventional beam-flexure methods lies in the fact that the specimen (graphene flake or graphite crystal) can be located at any point along the flexed span and not just at the centre. Thus, simultaneous studies on multiple spots (specimens) can be performed on the same beam. Furthermore, the experimental arrangement described in the Experimental section, allows us to reverse the direction of flexure and to conduct compression measurements as well[19]. Finally, plastic substrates cannot be easily polished to nanometre flatness and the presence of impurities, grease or even additives may significantly reduce the strength of the van der Waals forces between exfoliated graphene and polymer. To avoid slippage, we have conducted parallel measurements on a graphene flake placed on the substrate and one embedded within the PMMA bar. For reference, we have also monitored simultaneously, the variation of the two components of the 2D peak, $2D_1$ and $2D_2$, in bulk graphite.

**Fig.1** sketches the experimental set-up, with the two cantilever beams for the "bare" and embedded "specimens", respectively. The top surface of the beam can be subjected to a gradient of applied strain by flexing it by means of an adjustable screw at the edge of the beam span. The maximum deflection of the neutral axis of the beam (elastic behaviour), is given by the following equation (see Experimental):

$$\varepsilon(x) = \frac{3t\delta}{2L^2}\left(1 - \frac{x}{L}\right) \quad (1)$$

where $L$ is the cantilever beam span, $\delta$ is the deflection of the beam (at the free end) at each increment of flexure and $t$ is the beam thickness. The position where Raman measurements are taken is denoted by the variable "*x*". For the above equation to be valid, the span to maximum deflection aspect ratio should be greater than 10.[20]

In Fig. 2 plots the Raman spectra taken from the graphene flakes in "bare" (Fig.2a) and "embedded" configuration (Fig.2b). As can be seen from the corresponding micrographs the flake is invisible in the "bare" configuration but it can be discerned in the "embedded" configuration due to the presence of the SU8 interlayer[13],[21], also see Experimental section. The sharp and symmetric 2D peak at 2680 cm$^{-1}$ is the Raman fingerprint of graphene.[10] For comparison, Fig. 2 also shows the Raman spectrum from an adjacent graphite crystal, with the characteristic doubling of the 2D peak.[10], [22] We note that for the "embedded" graphene, a clear 2D peak can be seen through 100 nm thick PMMA. This shows the feasibility of monitoring by means of Raman microscopy graphene materials incorporated in transparent polymer matrices, which are now the focus of intense research.[23] The relationship between Raman shift and strain (or stress) also means that in graphene/polymer nanocomposites, the reinforcement (i.e. the incorporated graphene) can also act as the material mechanical sensor. This has already put in good use in carbon fibre/ polymer composites and has served to resolve the role of the interface in efficient stress transfer[24] and the fracture processes in unidirectional[25], but also multidirectional composites.[26]

Fig. 3a plots the fitted Position of the 2D peak as a function of strain for a monolayer graphene, Pos(2D), and bulk graphite, Pos(2D$_1$), Pos(2D$_2$), laid out on the PMMA substrate. In tension, Pos(2D) decreases with strain. A simple fitting given for example by a second-degree polynomial captures fairly well- within experimental error- the observed trend. The right-hand side $\partial\omega_{2D}/\partial\varepsilon$ axis measures the first derivative of the fit, which is a straight line that ranges from -10 cm$^{-1}$ / % near the origin to a maximum of ~ -60 cm$^{-1}$/ % at 0.9% strain. Indeed forcing a straight line to the data may underestimate the value of Raman shift rate, particularly if the experiment terminates at low strains. On the other hand the results in compression are quite different. $\partial\omega_{2D}/\partial\varepsilon$ seems to diminish from an initial value of +25 cm$^{-1}$ / % to zero at 0.74% compressive strain. The unsmooth transition through the zero point is an indication of the presence of residual strain in the material at rest. This could be the result of the placement process and the induced changes in graphene topology on the given substrate (see Fig. 1). The deposited flake interacts by van der Waals forces with the substrate but is "bare" on the outer surface. Hence, it is not surprising that under these conditions a compressive force would gradually detach the flake from the substrate, as manifested by the much lower initial slope in compression and the subsequent plateau at high strains. Finally, the graphite flake placed on top of the PMMA seems to be loaded only marginally upon the application of tensile load (Fig.3b). Again, this is to be expected since the weak forces that keep the crystal attached to the substrate are not sufficient to allow efficient stress transfer through the thickness of the whole graphitic block.

Fig.4 shows the results for the embedded sample. Here, the graphene is fully surrounded by polymer molecules and the stress transfer is far more efficient upon flexure of the beam. However, the initial drag in the 2D peak shift in tension and the sudden uptake observed in compression indicate that the flake is again under a residual compressive strain. This strain might also originate from the treatment of the top PMMA layer (Fig 1b), which might shrink during drying. When subjected to tension, a certain deformation will be needed to offset the initial compression and then a significant decrease of Pos(2D) is observed. However, the unfolding of the intrinsic ripples[27] of the stable graphene could also play a part since the parabolic fit to the data seems to hold satisfactorily up to 1.3% (see Fig. 4a). In other words, when a rippled material (equilibrium condition) is stretched, there will be a point in the deformation history whereby a greater portion of the mechanical energy will contribute to bond stretching rather than the "unfolding" of the structure. In compression, the sudden increase of Pos(2D) upon loading is an outcome of (a) the efficient stress transfer due to the incorporation of the material into the substrate and (b) the flake being already under compression at rest. Again a second-order polynomial captures fairly well the observed trend. The observed $\partial\omega_{2D}/\partial\varepsilon$ in compression is ~+59 cm$^{-1}$/ % near the origin (assuming absolute values of strain, see Experimental), which is similar to the maximum shift in tension, again confirming the presence of residual strain of compressive nature at rest. We note that these values are in excellent agreement with our previous tensile measurements on "bare" graphene done at extremely small strain rates.[13] Note that at -0.6% strain, the flake starts collapsing in compression as manifested by the inflection of Pos(2D) vs strain curve (Fig. 4a) and the subsequent relaxation of the Raman shift values.[19]

The classical theory of elasticity requires that since the thickness of a graphene monolayer is essentially zero then the flexural rigidity should also be zero. However, atomistic scale simulations[28],[29] predict that the bond-angle effect on the interatomic interactions should result in a finite flexural rigidity defined in each case by the interatomic potential used. The tension rigidity, C, of graphene at the unstrained equilibrium state for uniaxial stretching and curvature as derived by atomistic modelling[29] is given by:

$$C = \frac{1}{2\sqrt{3}}\left[\left(\frac{\partial^2 V_{ij}}{\partial r_{ij}^2}\right)_0 + \frac{B}{8}\right] \quad (2)$$

and the flexural rigidity, D, by

$$D = \frac{\sqrt{3}}{4}\left(\frac{\partial V_{ij}}{\partial \cos\theta_{ijk}}\right)_0 \quad (3)$$

where $V_{ij}$ is the interatomic potential and $\theta_{ijk}$ is the angle between two atomic bonds i-j and i-k (k≠i, j), $r_{ij}$ is the length of the bonds and *B* is an expression of the interatomic potential employed. The partial derivative of eq. (2) would be zero without the multibody coupling term as explained in [29].

The ratio of flexural to tension rigidities for uniaxial tension and bending is given by:

$$\frac{D}{C} = \frac{h^2}{12} \qquad (4)$$

where $h$ is the thickness of the plate/ shell. Finally, the critical strain, $\varepsilon_c$, for the buckling of a rectangular thin shell under uniaxial compression is given by:[20]

$$\varepsilon_c = \frac{\pi^2 k D}{C w^2} \qquad (5)$$

where $w$ is the width of the flake and $k$ is a geometric term.

The dimensions of the graphene monolayer used in the experiment were approximately 30 μm width and 100 μm in length (k=3.6). The tension rigidity (equation (2)) predicted by atomistic modelling using Brenner (2002) potentials[30] for both zigzag and armchair nanotubes at zero curvature that approximates graphene is comparable to 340 GPa nm which is the value measured recently in graphene by AFM.[4] Using this value we can derive from equation (4), the flexural rigidity of free graphene to be 3.18 GPa nm$^3$. Henceforth, the critical buckling strain for a flake of w=30 μm can now be calculated from equation (5) to be 300 microstrain or -0.03%. This indicates that a free graphene will collapse (buckle) at rather small axial compressive strains.

The experimental results presented here for an embedded graphene flake are very revealing. Firstly, as mentioned above, the Raman slope of about +59 cm$^{-1}$ /% measured at strains close to zero (very onset of the experiment) confirms that the flake can support fully in compression the transmitted load. However, the linear decrease of the Raman slope for higher strains up to about -0.7% is indicative of the gradual collapse of the material although it is still capable of supporting a significant portion of compressive load. It seems therefore that the graphene is prevented from full buckling by the lateral support offered by the surrounding material but at strains >-0.7% the interface between graphene and polymer possibly weakens or fails and the flake starts to buckle as it would do in air at -0.03%. The use of harder matrices or stronger interfaces between graphene/polymer matrix, should shift the critical strain for buckling to much higher values. The fact that one-atom-thick monolayers embedded in polymers can provide reinforcement in compression to high values of strain (in structural terms) is the most significant outcome of this work and paves the way for the development of nanocomposites for structural applications. It is interesting, however, to note that even the "bare" flake, which has only partial lateral support, can still be loaded axially in compression albeit at a less efficient rate than the embedded graphene. All the above is a very important area of future research and will serve to provide a link between nano and macro mechanics. For a purely elastic analysis, if we assume a graphene elastic modulus of 1 TPa [11], then the results presented here would be translated to an axial buckling stress of 6 GPa. This is at least three times higher than commercial CF in spite of the large diameters (7 µm) of CF and, hence, their higher Euler-instability threshold.[19]

Finally, for bulk graphite the results in Fig 4b show that by embedding the crystal in a thin layer of polymer a dramatic improvement in the stress transfer is obtained. To our knowledge this is the first time the 2D peak variation with tensile strain for bulk graphite is measured (see Supporting Material). In this case Pos(2D) changes linearly with strain, which again points to the fact that graphene layers in graphite are straight, as opposed to the "wrinkled" nature of the graphene monolayers.[27] Future work needs to assess the stress uptake of the atomic bonds in the whole range of graphitic materials from nanoscale graphene to macroscopic CF.

*Experimental section*

Graphene monolayers were prepared by mechanical cleavage from natural graphite (Nacional de Grafite) and transferred onto the PMMA cantilever beam. A sketch of the jig and the beam dimensions are shown in Fig.1. The beam containing the "bare" graphene/graphite specimens is composed solely of poly(methyl methacrylate) (PMMA) with thickness $t$ = 8.0 mm and width b = 10.0 m. The graphene flake is located at a distance, x, from the fixed end of 11.32 mm. The beam containing the "embedded" graphene/ graphite is made of a layer of PMMA and a layer of SU8 (~200 nm) photoresist of similar Young's modulus with thickness t = 2.9 mm and width b = 12.0 m. The graphene flake is located at a distance, x, from the fixed end of 10.44 mm. The SU8 also serves to increase the optical contrast.[13],[21] After placing the samples, another thin layer of PMMA (~100 nm) was laid on the top. The top surface of the beam can be subjected to a gradient of applied strain by flexing the beam by means of an adjustable screw positioned at a distance L = 70.0 mm from the fixed end (Fig.1). The deflection of the neutral axis of the beam (elastic behaviour), is given by:[20]

$$\delta = \frac{PL^3}{3EI} \qquad (6)$$

where P is the concentrated load applied to the end of the beam, L is the span of the beam, E is the Young's modulus of the beam material and I is the moment of inertia of the beam cross-section. The deflection δ was measured accurately using a dial gauge micrometer attached to the top surface of the beam. The mechanical strain as a function of the location (x,y) is given by:[20]

$$\varepsilon(x,y) = \frac{M(x) \cdot y}{EI} \qquad (7)$$

where M(x) is the bending moment along the beam, x the horizontal coordinate (distance from fixed end) and y the vertical coordinate (distance from neutral axis). In our case, the mechanical strain at the top surface of the beam (i.e. y=t/2) and, hence on a fixed graphene/ graphite position, is given by:

$$\varepsilon\left(x,\frac{t}{2}\right) = \frac{PLt\left(1-\frac{x}{L}\right)}{2EI} \qquad (8)$$

By substituting equation (6) into (8), the strain as a function of the position $x$ along the beam span and on the top surface of the beam (equation (1)) is derived. The validity of this method for measuring strains within the -1.5% to +1.5% strain range has been verified earlier.[19]

Raman spectra are measured at 514.5 nm (2.41eV) with a laser power of below 1 mW on the sample to avoid laser induced local heating. A 100x objective with numerical aperture of 0.95 is used, and the spot size is estimated to be ~1 µm. The data are collected in back-scattering and with a triple monochromator and a Peltier cooled CCD detector system. The spectral resolution is ~2cm$^{-1}$. The polarization of the incident light was kept parallel to the applied strain axis. Raman spectra of both graphene and graphite were fitted with Lorentzians. The full width at half maxima (FWHM) for the unstressed graphene was found to be approximately 27 cm$^{-1}$. No significant differences in FWHM between "bare" and "embedded" flakes were detected. The FWHM increased with strain in tension for both "bare" and "embedded" flakes; a maximum increase by 10 cm$^{-1}$ were measured at approximately 0.9% for both cases. However, in compression a similar increase was only noted in the case of the "embedded" flake whereas the FWHM in the case of the "bare" specimen seems to be fluctuating around the initial value at zero strain.

Fig. 2 shows some representative Raman spectra of the 2D band of bulk graphite and the characteristic double structure is evident.[22] The most intense peak, $2D_2$, is located at ~2730 cm$^{-1}$ and the weaker one, $2D_1$, at ~2690 cm$^{-1}$. The application of mechanical tension shifts both components towards lower frequencies at similar rates, as for Fig. 3 and Table 1. Close inspection of the Raman spectra obtained from different points of the graphene flakes shows a non-uniform strain distribution. Strain evolution in both samples was followed in the vicinity of points exhibiting 2D peak position at ~2690cm$^{-1}$ at zero strain. The error bars in Figs 3 and 4 correspond to the standard deviation of at least 5 spectra taken from spots around these reference points. Loading and unloading experiments showed no hysteresis within the range of strains applied here. Finally, for the data fittings in compression, absolute values of strain were used in order to show positive values of slope in compression; this is in agreement with the convention used in the experiments involving hydrostatic pressure.[6]. However, in mathematical terms the strain in compression is considered as "negative" strain and since the variation in 2D peak position is positive, $\partial\omega_{2D}/\partial\varepsilon$ should also be negative up to the inflection point.

**Acknowledgements**

One of the authors (CG) would like to thank Prof. N. Melanitis (HNA, Greece) for useful discussions during the preparation of this manuscript. FORTH/ ICE-HT acknowledge financial support from the Marie-Curie Transfer of Knowledge program CNTCOMP [Contract No.: MTKD-CT-2005-029876]. Also, GT gratefully acknowledges FORTH/ ICE-HT for a scholarship and ACF, KN, AKG thank the Royal Society and the European Research Council for financial support.

**Table 1** The values of wavenumber shift per % strain for the 2D line reported for various graphitic materials (excluding CNT)

| Reference | Maximum Strain Sensitivity ($cm^{-1}$/ %) for the 2D line in tension | | |
|---|---|---|---|
| | Graphene | Graphite | Carbon Fibres |
| Ni et al (2008) and Yu et al (2008) | -27.8* | - | - |
| Huang et al (2008) | -21.0* | - | - |
| Mohiuddin et al (2008) | ~-64* | - | - |
| Galiotis & Batchelder (1988) | - | - | -25 |
| This work | -59.1* | -1.3/ -2.1* | - |
| | +25.8 (compression)* | | |
| | -65.9** | -49.0/ -51.0** | |
| | +59.1 (compression)** | | |

*"Bare" graphene flake or graphite crystal on plastic substrate. For the work reported here, the graphene value was taken at 0.9% strain (Fig.3a).

**"Embedded" graphene flake or graphite crystal within the plastic substrate. The values in tension were taken at 1.3% strain and in compression near the origin (Fig.4 a, b). For the graphite the slopes correspond to the 2690 $cm^{-1}$ ($2D_1$) and 2730 $cm^{-1}$ ($2D_2$) bands, respectively.

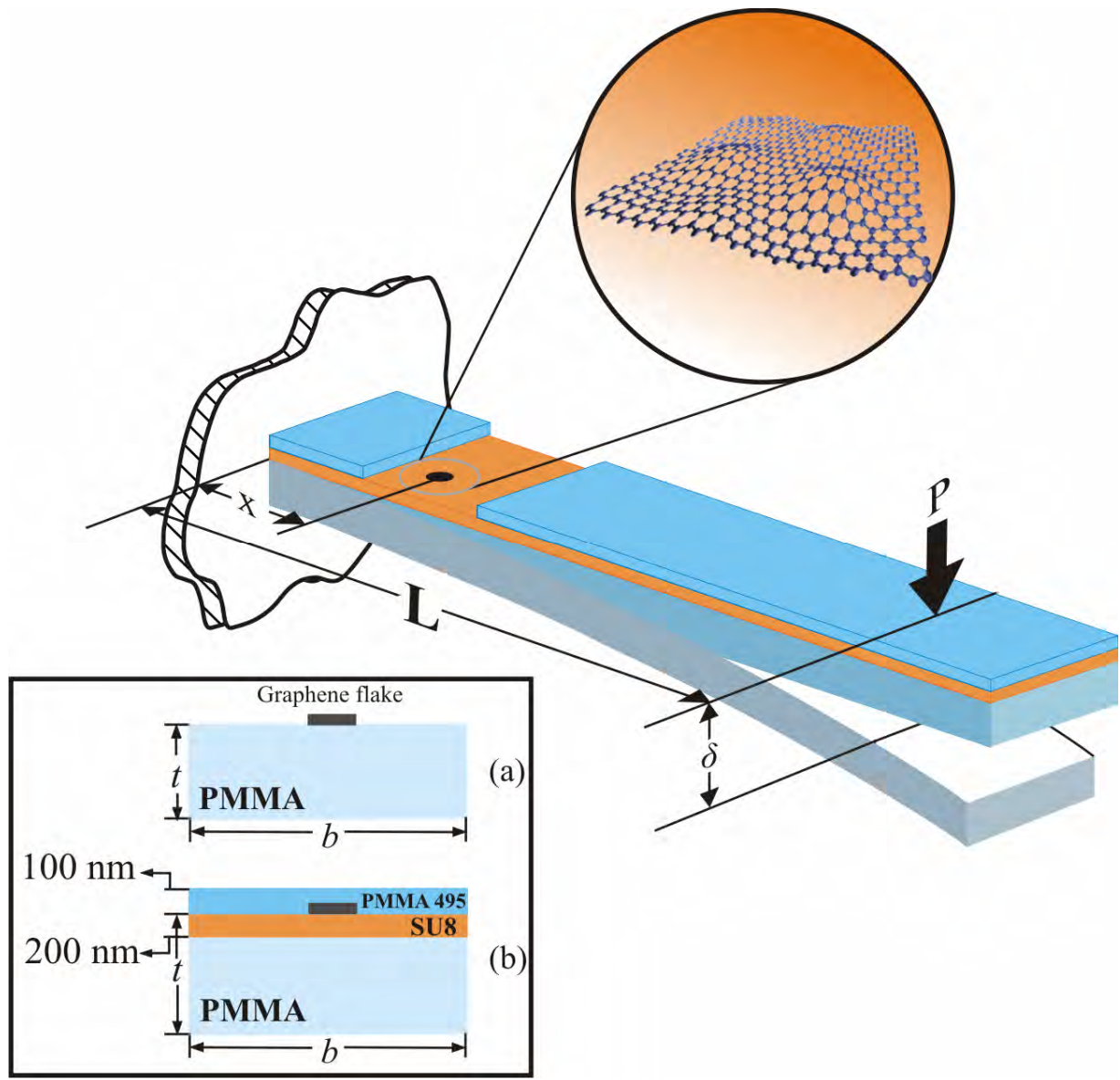

**Figure 1.** Cantilever beams for (a) "bare" and (b) "embedded" graphene flakes. The wrinkled nature of the graphene flake is portrayed.

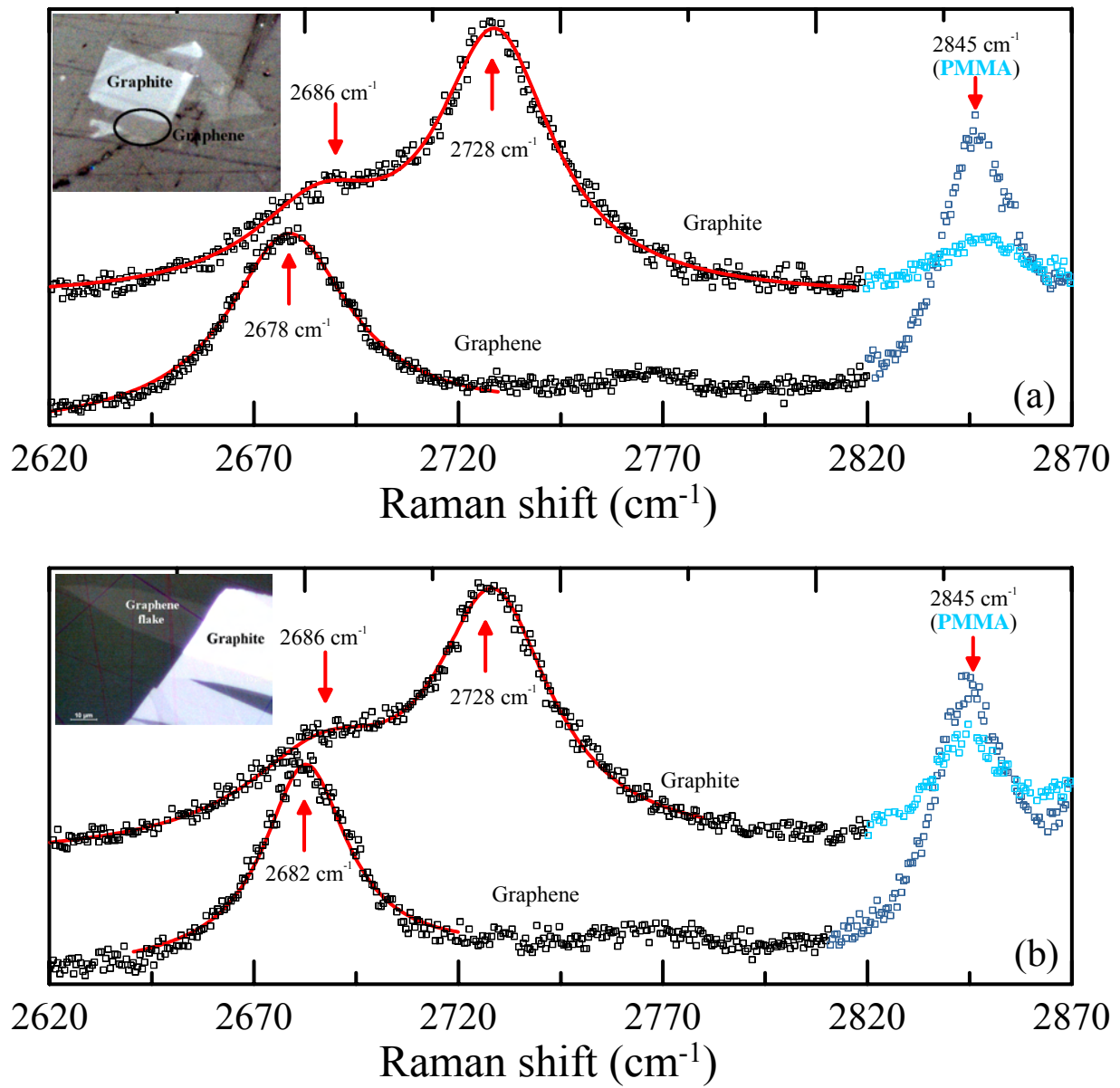

**Figure 2.** Sampling and corresponding Raman spectra from (a) "bare" and (b) "embedded" graphene flakes. Graphite spectra for each specimen are also shown. In all cases the PMMA Raman band at 2845 cm-1 is also shown. The observed ripples in the case (b) correspond to tiny striations in the upper PMMA film. The red lines represent Lorentzian fits to the graphene or graphite spectra.

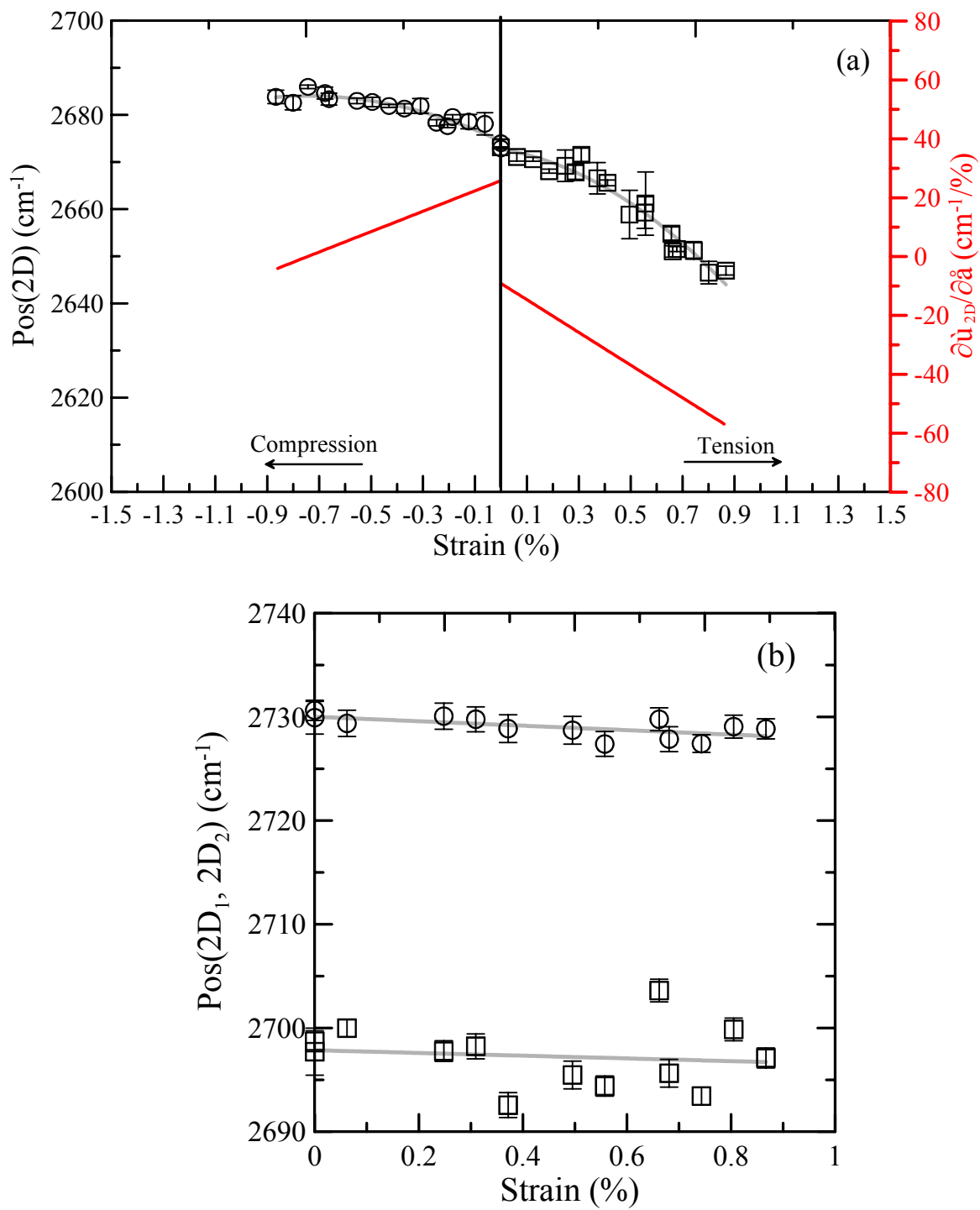

**Figure 3.** Wavenumber vs strain in both tension and compression for the (a) "bare" flake and (b) corresponding bulk graphite in tension. The second degree polynomial curves are of the form $\omega = 2672.8 - 9.1\varepsilon - 27.8\varepsilon^2$ and $\omega = 2674.4 + 25.8|\varepsilon| - 17.3\varepsilon^2$ for graphene in tension and compression, respectively. For the graphite in tension the results were least-squares-fitted with a straight line of slope of -1.3 and -2.1 cm$^{-1}$/% for peaks at 2690 cm$^{-1}$ (2D$_1$) and 2730 cm$^{-1}$ (2D$_2$), respectively.

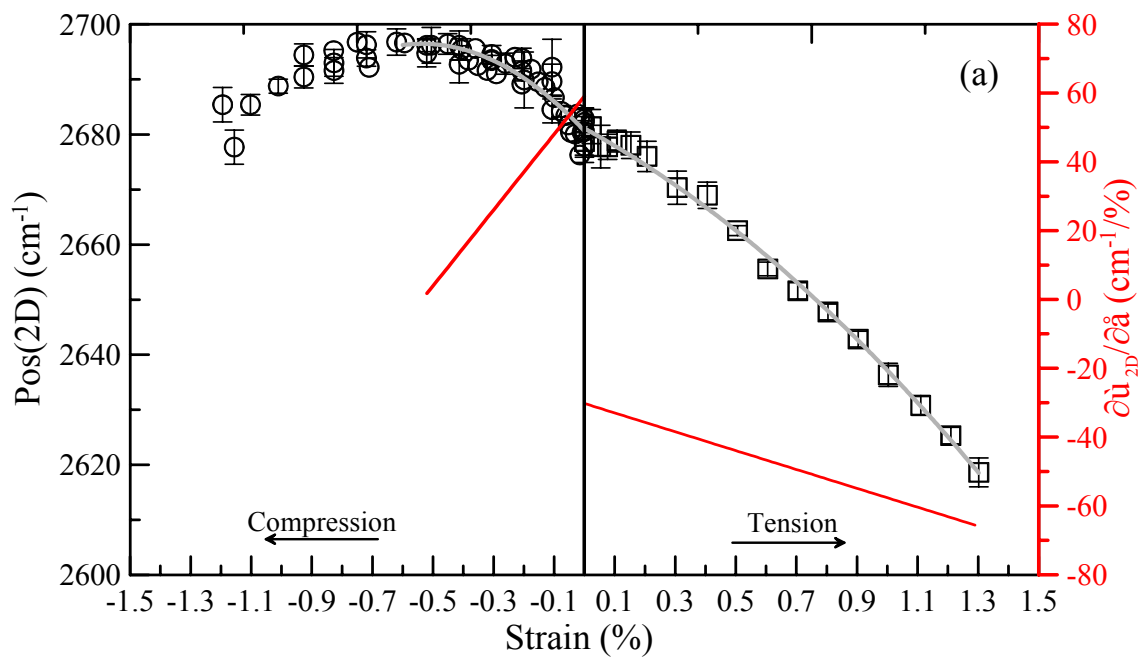

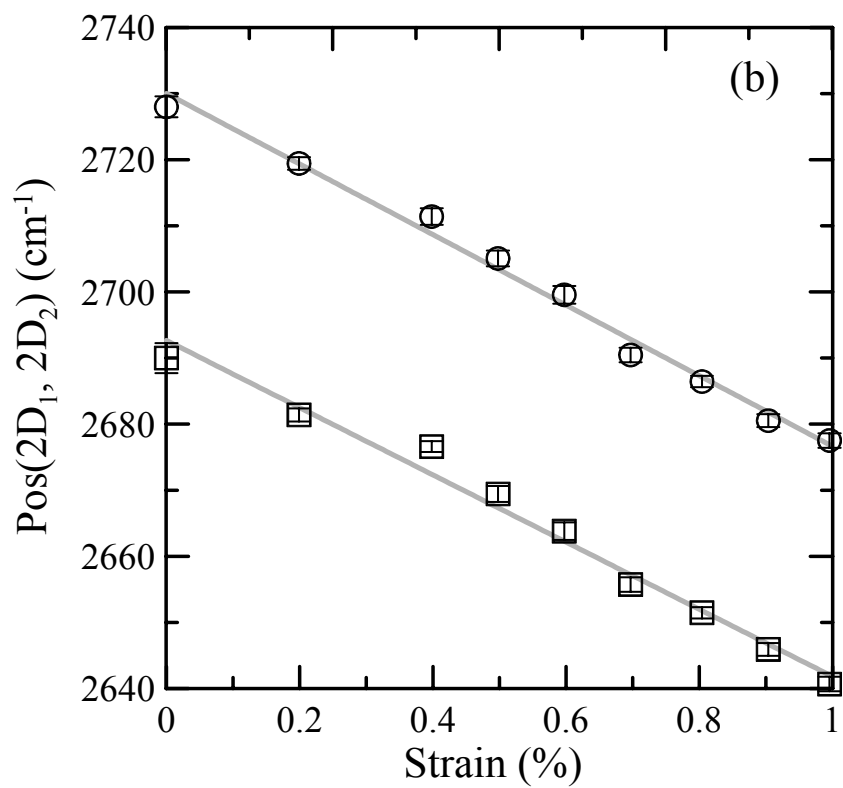

**Figure 4.** Wavenumber vs strain in both tension and compression for the "embedded" flake and corresponding bulk graphite in tension. The second degree polynomial curves are of the form $\omega = 2681.1 - 30.2\varepsilon - 13.7\varepsilon^2$ and $\omega = 2680.6 + 59.1|\varepsilon| - 55.1\varepsilon^2$ for graphene in tension and compression, respectively. For the graphite in tension the results were least-squares-fitted with a straight line of slope -50.9 cm$^{-1}$/% for the 2730 cm$^{-1}$ peak and 53.4 cm$^{-1}$/% for the 2693 cm$^{-1}$ peak.

# Supporting Information

## Subjecting a graphene monolayer to tension and compression

*Georgia Tsoukleri, John Parthenios, Konstantinos Papagelis, Rashid Jalil, Andrea C. Ferrari, Andre K. Geim, Kostya S. Novoselov and Costas Galiotis*

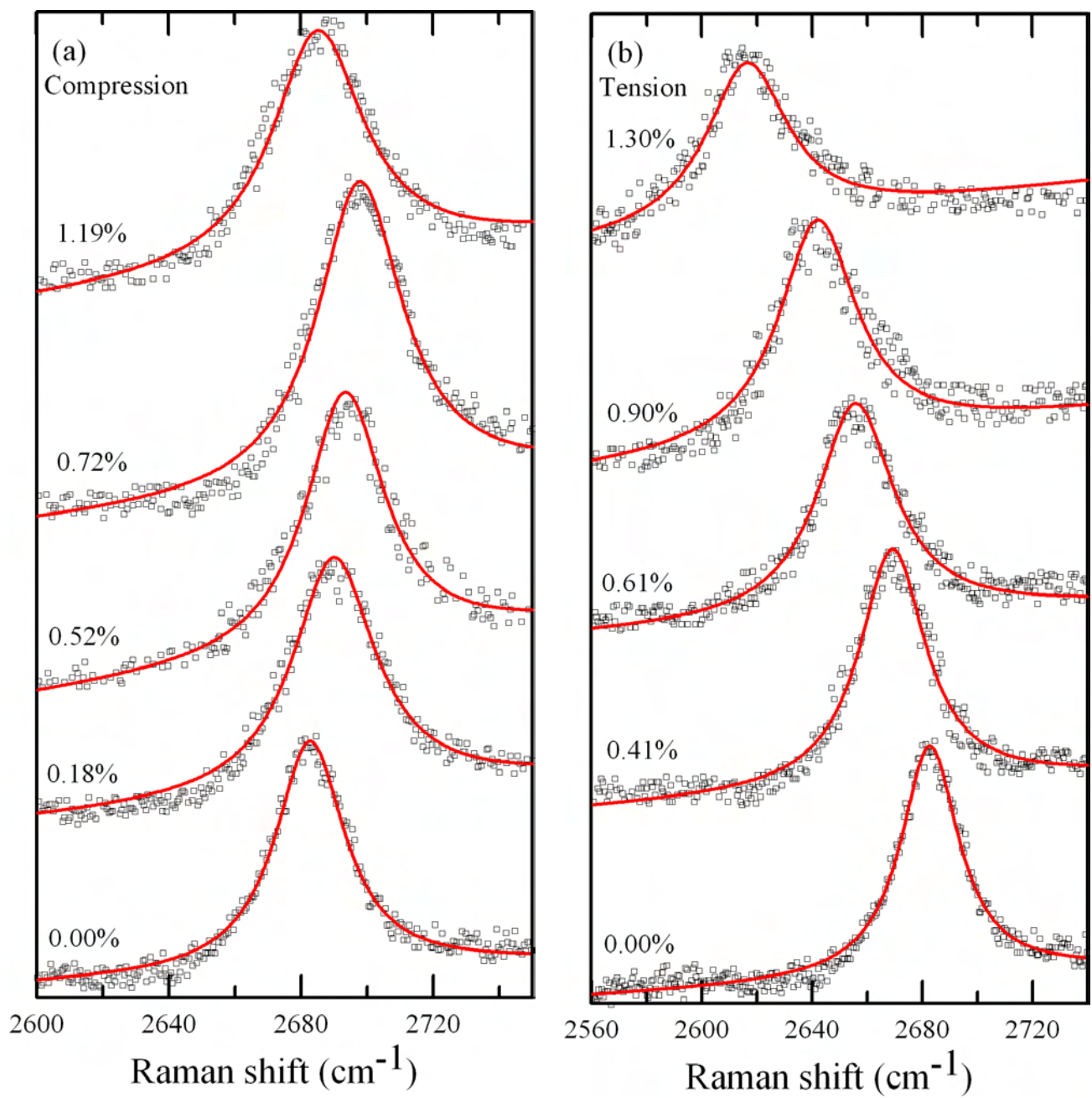

**Supporting fig. 1** 2D peak of the "embedded" graphene flake (a) in compression and (b) in tension as a function of uniaxial strain. The strains are indicated on the left side of the spectra. Red lines represent Lorentzian fits.

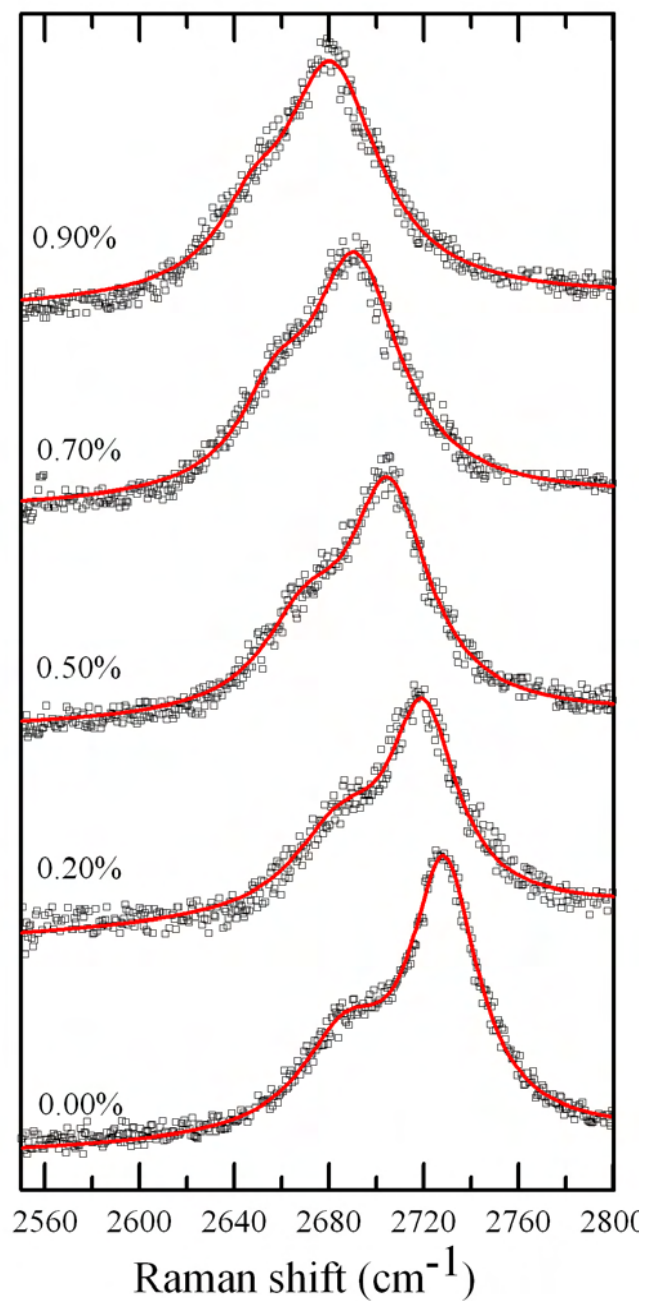

**Supporting fig. 2** 2D peak of bulk graphite of the "embedded" graphene flake as a function of uniaxial strain. The strains are indicated on the left side of the spectra. The red lines are fitted curves consisted of two Lorentzian components corresponding to $2D_1$ and $2D_2$ peaks.

**Supporting table 1** Summary of fitting parameters* (in ref. Figs 3 and 4)

|  | "Bare" flake specimen | | | | "Embedded" flake specimen | | | |
| --- | --- | --- | --- | --- | --- | --- | --- | --- |
|  | Graphene | | Graphite | | Graphene | | Graphite | |
|  | Tension | Compression | $2D_1$ | $2D_2$ | Tension | Compression | $2D_1$ | $2D_2$ |
| $\alpha_0$ (cm$^{-1}$) | 2672.8 | 2674.4 | 2697.9 | 2730.0 | 2681.1 | 2680.6 | 2692.7 | 2730.0 |
| $\alpha_1$ (cm$^{-1}$/%) | -9.1 | 25.8 | -1.3 | -2.1 | -30.2 | 59.1 | -53.4 | -50.9 |
| $\alpha_2$ (cm$^{-1}$/%$^2$) | -27.8 | -17.3 | 0.0 | 0.0 | -13.7 | -55.1 | 0.0 | 0.0 |

*Fitting function: $\Delta\nu(\varepsilon) = \alpha_0 + \alpha_1 |\varepsilon| + \alpha_2 \varepsilon^2$